# Aggression in the workplace makes social distance difficult

Keisuke Kokubun[1]

1 Economic Research Institute, Japan Society for the Promotion of Machine Industry

## Abstract

The spread of new coronavirus (COVID-19) infections continues to increase. The practice of social distance attracts attention as a measure to prevent the spread of infection, but it is difficult for some occupations. Therefore, in previous studies, the scale of factors that determine social distance has been developed. However, it was not clear how to select the items among them, and it seemed to be somewhat arbitrary. In response to this trend, this paper extracted eight scales by performing exploratory factor analysis based on certain rules while eliminating arbitrariness as much as possible. They were Adverse Conditions, Leadership, Information Processing, Response to Aggression, Mechanical Movement, Autonomy, Communication with the Outside, and Horizontal Teamwork. Of these, Adverse Conditions, Response to Aggression, and Horizontal Teamwork had a positive correlation with Physical Proximity, and Information Processing, Mechanical Movement, Autonomy, and Communication with the Outside had a negative correlation with Physical Proximity. Furthermore, as a result of multiple regression analysis, it was shown that Response to Aggression, not the mere teamwork assumed in previous studies, had the greatest influence on Physical Proximity.

## Introduction

The spread of new coronavirus (COVID-19) infections is increasing steadily. As of this writing, August 8, 2020, there are 19,382,107 infected people worldwide (Johns Hopkins University & Medicine, 2020). The recommended practice of preventing infection is the social distance, that is, maintaining the distance between people (World Health Organization, 2020). However, some jobs are easy and difficult to take social distance. Therefore, research to clarify the relationship between work characteristics and social distance is becoming active (Crowley & Doran, 2020; Dingel & Neiman, 2020; Hatayama et al., 2020; Koren & Pető, 2020). Such research can give great hints and significance to the new industrial structure in the era of with-corona. However, it does not mean that there is no problem at all, such as making people feel arbitrary about how to make the scale.

Therefore, this paper uses 98 items recorded in the Work Context and Work Activities of O*NET to perform factor analysis while eliminating arbitrariness as much as possible. Then, after creating an independent variable from the extracted factors, a regression analysis using Physical

Proximity as the dependent variable is performed. The purpose of this is to find the factors that influence the social distance (Physical Proximity as a proxy variable). At the same time, Koren & Pető (2020)'s Teamwork, Customer and Presence, and Dingel & Neiman (2020)'s Remote Working Index will be used as independent variables to verify whether these are useful as determinants of social distance.

## Review of previous studies

Research to clarify the relationship between work characteristics and social distance is becoming more active (Crowley & Doran, 2020; Dingel & Neiman, 2020; Hatayama et al., 2020; Koren & Pető, 2020). Among them, the Social Distance Index developed by Koren & Pető (2020) and the Remote Working Index developed by Dingel & Neiman (2020) have 49 and 266 citations respectively (as of August 8, 2020, computed by Google Scholar). However, there is some arbitrariness in how to create the indicators. The former classifies 14 questionnaire items recorded in O*NET, an occupational information site in the United States, into these 3 categories, based on the hypothesis that three factors that influence social distance are Teamwork, Customer, and Presence. The latter also creates variables on the assumption that 17 items related to outdoor work etc. recorded in O*NET influence the executability of remote work. However, neither of them clearly shows the criteria for classification. Probably they are classified by the authors' eyes.

However, there is no guarantee that the respondents to the questionnaire have responded as the researchers intended. For example, the variables created as Teamwork by Koren & Pető (2020) include items related to leadership. Perhaps there is a difference in the possibility of social distance between vertical and horizontal teams. In the first place, it is a problem that the correlation between Work Context - Physical Proximity, the items seemingly closest to the feasibility of social distance and remote work among the items recorded in O*NET, and each scale is not verified.

Therefore, in this paper, to eliminate arbitrariness as much as possible, some factors common to various occupations will be extracted by performing exploratory factor analysis based on certain rules. And based on those factors, we will show the factors that determine Physical Proximity by performing multiple regression analysis using the constructed independent variables. At the same time, we will analyze Teamwork, Customer, Presence by Koren & Pető (2020), and the Remote Working Index by Dingel & Neiman (2020) as independent variables, and verify whether these are useful as determinants of social distance.

## Method

Among questionnaire results for 968 occupations posted on the O*NET website (https://www.onetonline.org/), we use the 98 items recorded in Work Context and Work Activities which were used in the studies of Koren & Pető (2020), Dingel & Neiman (2020) and others.

Importance and Level are recorded in Work Activities, but Importance is used in the current research following previous research. All items are in the format of making a number from 0 to 100 for frequency and importance. The criterion for factor extraction is eigenvalue 1 or more, and the factor load is calculated after performing varimax rotation by the main factor method. After that, items with a factor load of less than 0.4 and items with a factor load of 0.4 or more on a plurality of factors are excluded, and factor analysis is performed again using the same criteria. After that, this process is repeated until there are no items whose factor loads are less than 0.4 and items whose factor loads are 0.4 or more. Here, we follow the idea of Stevens (1992) who suggests using a cut-off of 0.4, irrespective of sample size, for interpretative purposes. After the factor structure is established, this paper presents a simple hypothesis that should be verified and performs a regression analysis using the variables consisting of each factor as the independent variable and Physical Proximity as the dependent variable to analyze the factors that influence social distance.

As a result of repeating the factor analysis 6 times by the above method, as shown in Table 1, eight factors consisting of 46 items were extracted. Note that the sentences listed are not the question text, but the content of the question text. Details of questions and options can be referred to on O*NET, so they are omitted in this article. However, to give an example, the question sentence corresponding to the content sentence at the top of the table, "Work Context - Very Hot or Cold Temperatures" is "How often does this job require working in very hot (above 90 F degrees) or very cold (below 32 F degrees) temperatures?", and the options are "Never", "Once a year or more but not every month", "Once a month or more but not every week", "Once a week or more but not every day", and "Every day". For each answer, 0 points, 25 points, 50 points, 75 points, and 100 points are assigned, and the average value is calculated for each occupation. Also, the Physical Proximity used as a dependent variable in the current research is selected from "I don't work near other people (beyond 100ft.)", "I work with others but not closely (e.g., private office)", "Slightly close (e.g., shared office)", "Moderately close (at arm's length)", and "Very close (near touching)" for the question sentence of "To what extent does this job require the worker to perform job tasks in close physical proximity to other people?" and is similarly assigned a value of 0 to 100 when totaling.

Based on the contents of the included items, the factors were named as Adverse Conditions, Leadership, Information Processing, Response to Aggression, Mechanical Movement, Autonomy, Communication with the Outside, and Horizontal Teamwork. By the way, Physical Proximity is not included in any of these variables because it loaded multiple factors in the first factor analysis. As indicated by the symbol (R) at the end of the sentence, four of Adverse Conditions overlap with Remote Working in Dingel & Neiman (2020). Also, as shown in (P), one of them overlaps with the Presence of Koren & Pető (2020). Similarly, three of the Leadership and one of Horizontal Teamwork overlap with Teamwork of Koren & Pető (2020) as shown in (T). One of Response to Aggression overlaps with the Presence of Koren & Pető (2020) as shown in (P). There were no duplicate items with Customer of Koren & Pető (2020). Overall, 9 of 46 items, or 19.6%, overlapped with either or both of Dingel & Neiman (2020) and Koren & Pető (2020).

Table 1 Results of exploratory factor analysis

| Item | Adverse Conditions | Leadership | Information Processing | Response to Aggression | Mechanical Movement | Autonomy | Communication with the Outside | Horizontal Teamwork |
|---|---|---|---|---|---|---|---|---|
| Work Context - Very Hot or Cold Temperatures | ***0.916*** | 0.003 | -0.110 | 0.031 | 0.005 | -0.101 | 0.007 | -0.127 |
| Work Context - Extremely Bright or Inadequate Lighting | ***0.880*** | -0.086 | -0.018 | 0.102 | 0.054 | -0.047 | -0.040 | 0.043 |
| Work Activities - Operating Vehicles, Mechanized Devices, or Equipment (P) (R) | ***0.863*** | 0.025 | -0.042 | 0.090 | 0.054 | -0.024 | -0.020 | -0.189 |
| Work Context - Outdoors, Exposed to Weather (R) | ***0.857*** | 0.035 | 0.008 | 0.116 | -0.158 | 0.023 | 0.328 | -0.058 |
| Work Context - Exposed to Hazardous Equipment | ***0.851*** | -0.011 | -0.111 | -0.124 | 0.197 | -0.011 | -0.261 | -0.052 |
| Work Context - Exposed to High Places | ***0.834*** | -0.007 | -0.033 | -0.102 | 0.023 | -0.023 | 0.033 | 0.273 |
| Work Context - Indoors, Not Environmentally Controlled | ***0.833*** | 0.049 | -0.043 | -0.090 | 0.021 | 0.002 | 0.014 | -0.054 |
| Work Context - In an Open Vehicle or Equipment | ***0.824*** | 0.016 | -0.080 | -0.075 | -0.015 | -0.044 | -0.005 | -0.032 |
| Work Context - Spend Time Climbing Ladders, Scaffolds, or Poles | ***0.787*** | -0.007 | -0.079 | -0.133 | -0.028 | -0.034 | 0.023 | 0.262 |
| Work Context - Exposed to Minor Burns, Cuts, Bites, or Stings (R) | ***0.775*** | -0.036 | -0.234 | 0.002 | 0.069 | -0.019 | -0.258 | -0.126 |
| Work Context - Outdoors, Under Cover (R) | ***0.774*** | 0.069 | 0.040 | 0.019 | -0.164 | 0.045 | 0.355 | 0.023 |
| Work Context - Exposed to Whole Body Vibration | ***0.761*** | -0.030 | -0.065 | -0.016 | -0.020 | -0.051 | -0.108 | 0.076 |
| Work Context - Sounds, Noise Levels Are Distracting or Uncomfortable | ***0.718*** | -0.060 | -0.128 | 0.126 | 0.272 | -0.141 | -0.299 | 0.070 |
| Work Context - Spend Time Keeping or Regaining Balance | ***0.705*** | -0.022 | -0.222 | 0.155 | -0.109 | -0.083 | -0.142 | 0.076 |
| Work Context - Indoors, Environmentally Controlled | ***-0.697*** | 0.124 | 0.237 | 0.041 | 0.126 | 0.113 | 0.125 | 0.171 |
| Work Context - Work Schedules | ***0.468*** | -0.004 | -0.123 | -0.099 | -0.284 | 0.066 | 0.159 | -0.090 |
| Work Activities - Guiding, Directing, and Motivating Subordinates (T) | 0.045 | ***0.894*** | 0.192 | 0.053 | -0.016 | 0.060 | -0.004 | 0.109 |
| Work Activities - Coordinating the Work and Activities of Others (T) | 0.070 | ***0.855*** | 0.188 | 0.073 | -0.014 | -0.013 | 0.066 | 0.194 |
| Work Activities - Staffing Organizational Units | -0.088 | ***0.833*** | 0.141 | 0.069 | 0.041 | 0.063 | 0.169 | -0.015 |
| Work Activities - Developing and Building Teams (T) | -0.057 | ***0.819*** | 0.262 | 0.169 | -0.058 | -0.067 | 0.061 | 0.207 |
| Work Activities - Coaching and Developing Others | -0.065 | ***0.796*** | 0.238 | 0.160 | -0.190 | 0.073 | -0.190 | 0.125 |
| Work Activities - Monitoring and Controlling Resources | 0.043 | ***0.733*** | 0.095 | -0.083 | 0.100 | 0.169 | 0.244 | -0.114 |
| Work Activities - Scheduling Work and Activities | -0.036 | ***0.678*** | 0.307 | 0.008 | -0.086 | 0.237 | 0.188 | -0.030 |
| Work Activities - Training and Teaching Others | -0.009 | ***0.624*** | 0.321 | 0.089 | -0.204 | 0.063 | -0.303 | 0.088 |
| Work Activities - Judging the Qualities of Things, Services, or People | 0.037 | ***0.539*** | 0.338 | 0.097 | 0.033 | 0.170 | -0.149 | -0.136 |
| Work Activities - Analyzing Data or Information | -0.260 | 0.257 | ***0.790*** | -0.179 | 0.068 | 0.113 | 0.102 | 0.051 |
| Work Activities - Processing Information | -0.248 | 0.209 | ***0.788*** | -0.098 | 0.261 | 0.006 | 0.104 | 0.002 |
| Work Activities - Interpreting the Meaning of Information for Others | -0.281 | 0.347 | ***0.759*** | 0.007 | -0.135 | 0.150 | 0.020 | 0.128 |
| Work Activities - Updating and Using Relevant Knowledge | -0.262 | 0.298 | ***0.741*** | -0.028 | -0.010 | 0.237 | 0.066 | 0.079 |
| Work Activities - Getting Information | -0.196 | 0.218 | ***0.724*** | 0.058 | 0.140 | 0.141 | 0.083 | 0.110 |
| Work Activities - Documenting/Recording Information | -0.208 | 0.211 | ***0.695*** | 0.189 | 0.147 | 0.120 | 0.117 | -0.002 |
| Work Activities - Identifying Objects, Actions, and Events | 0.139 | 0.240 | ***0.676*** | 0.104 | 0.109 | 0.088 | -0.139 | 0.016 |
| Work Activities - Evaluating Information to Determine Compliance with Standards | 0.095 | 0.314 | ***0.549*** | 0.155 | 0.332 | -0.093 | 0.050 | 0.070 |
| Work Context - Deal With Unpleasant or Angry People | -0.066 | 0.075 | -0.056 | ***0.886*** | 0.195 | -0.041 | 0.050 | 0.057 |
| Work Context - Deal With Physically Aggressive People (R) | 0.053 | 0.094 | 0.064 | ***0.817*** | -0.051 | -0.029 | -0.040 | -0.009 |
| Work Context - Frequency of Conflict Situations | -0.043 | 0.284 | 0.108 | ***0.738*** | 0.195 | 0.126 | 0.127 | 0.210 |
| Work Context - Importance of Repeating Same Tasks | -0.024 | -0.145 | 0.059 | 0.204 | ***0.694*** | -0.122 | 0.054 | 0.024 |
| Work Context - Importance of Being Exact or Accurate | -0.084 | -0.121 | 0.273 | -0.035 | ***0.686*** | 0.064 | -0.057 | 0.034 |
| Work Context - Degree of Automation | -0.053 | -0.049 | 0.026 | -0.004 | ***0.524*** | -0.270 | 0.080 | -0.019 |
| Work Context - Time Pressure | 0.081 | 0.072 | 0.076 | 0.062 | ***0.460*** | 0.088 | -0.041 | 0.089 |
| Work Context - Freedom to Make Decisions | -0.069 | 0.166 | 0.271 | 0.020 | -0.137 | ***0.815*** | 0.064 | 0.011 |
| Work Context - Structured versus Unstructured Work | -0.207 | 0.209 | 0.239 | -0.048 | -0.074 | ***0.772*** | 0.127 | 0.044 |
| Work Context – Telephone | -0.190 | 0.201 | 0.313 | 0.190 | 0.184 | 0.394 | ***0.597*** | 0.243 |
| Work Activities - Communicating with Persons Outside Organization | -0.253 | 0.350 | 0.399 | 0.169 | -0.091 | 0.250 | ***0.515*** | 0.031 |
| Work Context - Work With Work Group or Team (T) | -0.008 | 0.362 | 0.170 | 0.270 | 0.212 | -0.103 | 0.041 | ***0.496*** |
| Work Context - Face-to-Face Discussions | -0.004 | 0.251 | 0.223 | 0.156 | 0.154 | 0.223 | 0.069 | ***0.470*** |

Note(s): If the factor load is 0.4 or more, italic and bold type

(T) and (P) indicate that they overlap with Teamwork and Presence of Koren & Pető (2020).

(R) indicates overlapping with Remote Working of Dingel & Neiman (2020).

## Hypothesis

Of the eight factors, Leadership, Response to Aggression, and Horizontal Teamwork are factors that include relationships with people and are expected to require Physical Proximity. From this, the following hypotheses are derived.

H1: Leadership, Response to Aggression, Horizontal Teamwork have a positive correlation with Physical Proximity.

On the other hand, Information Processing, Autonomy, Mechanical Movement, and Communication with the Outside are considered to be jobs that do not require physical contact with people. From this, the following hypotheses are derived.

H2: Information Processing, Autonomy, Mechanical Movement, Communication with the Outside have a negative correlation with Physical Proximity.

Besides, Adverse Conditions does not generally require closeness to people, but it is expected that some people will be required to have closeness if the job requires more than one person in case of an unexpected situation. Therefore, the following hypothesis is derived.
H3: Adverse Conditions has a positive correlation with Physical Proximity.

What has the strongest correlation with Physical Proximity? Information Processing, Autonomy, Mechanical Movement, and Communication with the Outside are elements that do not require Physical Proximity, and at the same time, they are not elements that actively distance Physical Proximity. Therefore, it is expected that any of the Leadership, Response to Aggression, and Horizontal Teamwork, which express relationships with people, is most strongly correlated with Physical Proximity. Of these, leadership can be demonstrated over long distances depending on the content, so it is considered that the correlation with Physical Proximity is weakest. Virtually reconstructing the reporting line shouldn't be too difficult if the relationship between the top and bottom is clear. There is also a study showing that virtual is more effective than face to face for some leadership (Purvanova & Bono, 2009). In this regard, Horizontal Teamwork is an element that involves a more horizontal relationship, so it is considered that Physical Proximity is required compared to Leadership. This is because who takes the initiative depends on the tasks that change day by day. Also, when someone performs many tasks, adjustments are frequently made such that someone reduces the tasks. However, with the proper feedback and procedural justification from the boss, the adverse effects of role ambiguity can be significantly reduced (De Clercq & Belausteguigoitia, 2017; Jong, 2016). Therefore, it is expected that Response to Aggression has the greatest correlation with Physical Proximity. This is because collaboration with a person who is not good at emotional control is likely to induce an unexpected situation, so it is considered necessary to have him or her nearby and to cooperate with others around them. Previous studies have shown that maintaining physical proximity to students, for example, is most necessary for teachers to turn their negative consciousness into a positive one (den Brock et al., 2005). Therefore, the following hypothesis is derived.
H4: Response to Aggression has the strongest positive correlation with Physical Proximity.

## Analysis and result

Table 2 is descriptive statistics. In addition to the above eight variables, Physical Proximity, the three variables of Teamwork, Customer, and Presence by Koren & Pető (2020) and Remote Working by Dingel & Neiman (2020) are listed with their values and correlations by the author's calculation. In general, 0.7 or more is considered to be an acceptable reliability coefficient (Cortina, 1993), but several researchers have accepted it as 0.6 or more (Taber, 2018; van Griethuijsen et al., 2015). According to the latter criterion, all eight variables are acceptable. However, according to the former criteria, Mechanical Movement and Horizontal Teamwork are unacceptable. Therefore, for Mechanical Movement, we prepared a 3-item version without Work Context-Time Pressure, and for Horizontal

Teamwork, we prepared a 3-item version with Work Context-Coordinate or Lead Others, dropped in the first factor analysis because it was the same factor as the other two items but loaded on other factors, added. After confirming that the reliability coefficients of these variables were 0.7 or more, they were used for the following analysis alternatively (The result is omitted because it did not make a big difference).

Looking at the magnitude of the correlation coefficient, all except for the two categories of Leadership and Mechanical Movement show a statistically significant correlation with Physical Proximity at the 5% level. However, the values exceeding 0.1, which is the standard value for "small" by Cohen (1988), are observed in three variables of Response to Aggression ($r=0.456$, $p<0.01$), Autonomy ($r=-0.130$, $p<0.05$), Horizontal Teamwork ($r=0.306$, $p<0.01$). Of these, only two, Response to Aggression and Horizontal Teamwork, exceeded the standard value of 0.3 for the “medium” by Cohen (1988). By the way, as shown in the table, among the three variables of Koren & Pető (2020), it is Customer ($r=0.415$, $p<0.01$) and Presence ($r=0.197$, $p<0.01$) that exceed 0.1, and only Customer of these exceeded 0.3. Dingel & Neiman (2020)’s Remote Working ($r=-0.356$, $p<0.01$) also exceeded 0.3. However, the coefficient of Koren & Pető (2020)’s Teamwork ($r=0.098$, $p<0.01$) was below 0.1. This is probably because, as described at the beginning, items that represent horizontal teamwork and items that represent vertical teamwork are mixed in the variable.

Table 3 shows the results of regression analysis. Eight independent variables are individually inputted into the regression equation. Response to Aggression ($\beta=0.456$, $p<0.01$) and Horizontal Teamwork ($\beta=0.306$, $p<0.01$) showed a significant positive correlation, supporting H1. Information Processing ($\beta=-0.085$, $p<0.01$), Autonomy ($\beta=-0.129$, $p<0.01$), Communication with the Outside ($\beta=-0.065$, $p<0.05$) showed a significant negative correlation. This is the result of supporting H2. Furthermore, a significant positive correlation was shown in Adverse Conditions ($\beta=0.074$, $p<0.05$), supporting H3. However, regarding Leadership ($\beta = 0.057$, $p>0.05$) where a positive correlation was expected and Mechanical Movement ($\beta = 0.001$, $p>0.05$) where a negative correlation was expected, a significant correlation was not found at the 5% level. Comparing the adjusted R-squared, Response to Aggression was the largest at 0.207, followed by Horizontal Teamwork at 0.093 and Autonomy at 0.016. This shows that Response to Aggression has the greatest influence on social distance, and it can be said that this is the result of supporting H4.

Table 2 Descriptive statistics

| | Mean | SD | α | Physical Proximity | Adverse Conditions | Leadership | Information Processing | Response to Aggression | Mechanical Movement | Autonomy | Communication with the Outside | Horizontal Teamwork | [a]Teamwork | [a]Customer | [a]Presence |
|---|---|---|---|---|---|---|---|---|---|---|---|---|---|---|---|
| Physical Proximity | 60.300 | 16.875 | | | | | | | | | | | | | |
| Adverse Conditions | 18.542 | 17.538 | 0.929 | 0.074* | | | | | | | | | | | |
| Leadership | 48.440 | 11.806 | 0.939 | 0.056 | -0.058 | | | | | | | | | | |
| Information Processing | 66.133 | 11.940 | 0.923 | -0.085** | -0.307** | 0.553** | | | | | | | | | |
| Response to Aggression | 36.748 | 12.878 | 0.871 | 0.456** | -0.034 | 0.243** | 0.146** | | | | | | | | |
| Mechanical Movement | 58.955 | 9.699 | 0.694 | 0.001 | -0.038 | -0.083** | 0.195** | 0.186** | | | | | | | |
| Autonomy | 76.918 | 11.440 | 0.897 | -0.130** | -0.214** | 0.349** | 0.417** | 0.041 | -0.137** | | | | | | |
| Communication with the Outside | 67.569 | 18.633 | 0.810 | -0.064* | -0.314** | 0.447** | 0.585** | 0.312** | 0.084** | 0.534** | | | | | |
| Horizontal Teamwork | 84.353 | 8.904 | 0.656 | 0.306** | -0.052 | 0.438** | 0.392** | 0.373** | 0.188** | 0.155** | 0.385** | | | | |
| [a]Teamwork | 55.408 | 12.084 | 0.889 | 0.098** | -0.094** | 0.933** | 0.588** | 0.287** | -0.024 | 0.288** | 0.485** | 0.619** | | | |
| [a]Customer | 53.837 | 14.099 | 0.775 | 0.415** | -0.312** | 0.525** | 0.412** | 0.584** | -0.068* | 0.363** | 0.626** | 0.368** | 0.526** | | |
| [a]Presence | 38.202 | 19.354 | 0.911 | 0.197** | 0.814** | -0.040 | -0.258** | -0.064* | 0.082* | -0.272** | -0.403** | -0.060 | -0.107** | -0.259** | |
| [b]Remote Working | 57.767 | 13.832 | 0.914 | -0.356** | -0.838** | -0.067* | 0.176** | -0.147** | -0.042 | 0.177** | 0.255** | -0.061 | 0.011 | -0.029 | 0.920** |

Note(s): n = 968; *Significance at the 5% level; **Significance at the 1% level; [a] is from Koren & Pető (2020), [b] is from Dingel & Neiman (2020).

In addition, as shown in the table, all three variables by Koren & Pető (2020), Teamwork (β = 0.098, p <0.01), Customer (β = 0.414, p <0.01), and Presence (β = 0.198, p <0.01), also showed a significant correlation at the 1% level. However, among them, the adjusted R-squared of Customer was a significantly large value of 0.171. Dingel & Neiman (2020)'s Remote Working (β=-0.356, p<0.01) also showed a significant correlation at the 1% level, and its adjusted R-squared was relatively large at 0.126. Comparing the magnitude of the partial regression coefficient and the adjusted R-squared, the order is as follows: Response to Aggression (β=0.456, $R^2$=0.207), Customer (β=0.414, $R^2$=0.171), Remote Working (β=-0.356, $R^2$=0.126).

Table 3 Results of simple regression analysis with Physical Proximity as the dependent variable

| Variable | β | | $R^2$ | Adj-$R^2$ | F | |
|---|---|---|---|---|---|---|
| Adverse Conditions | 0.074 | * | 0.006 | 0.005 | 5.367 | * |
| Leadership | 0.057 | | 0.003 | 0.002 | 3.094 | |
| Information Processing | -0.085 | ** | 0.007 | 0.006 | 7.011 | ** |
| Response to Aggression | 0.456 | ** | 0.208 | 0.207 | 253.458 | ** |
| Mechanical Movement | 0.001 | | 0.000 | -0.001 | 0.001 | |
| Autonomy | -0.129 | ** | 0.017 | 0.016 | 16.369 | ** |
| Communication with the Outside | -0.065 | * | 0.004 | 0.003 | 4.056 | * |
| Horizontal Teamwork | 0.306 | ** | 0.094 | 0.093 | 99.691 | ** |
| [a]Teamwork | 0.098 | ** | 0.010 | 0.009 | 9.348 | ** |
| [a]Customer | 0.414 | ** | 0.172 | 0.171 | 200.134 | ** |
| [a]Presence | 0.198 | ** | 0.039 | 0.038 | 39.194 | ** |
| [b]Remote Working | -0.356 | ** | 0.127 | 0.126 | 139.189 | ** |

Note(s): n = 968; *Significance at the 5% level; **Significance at the 1% level; [a] by Koren & Pető (2020), [b] by Dingel & Neiman (2020).

The first column of Table 4 is the result of multiple regression analysis by the stepwise method. Response to Aggression (β=0.460, p<0.01), Horizontal Teamwork (β=0.288, p<0.01) showed a positive correlation and Information Processing (β=-0.096, p<0.01), Mechanical Movement (β=-0.099, p< 0.01) and Communication with the Outside (β=-0.254, p<0.01) showed a negative correlation. Among them, Mechanical Movement did not show a significant correlation by simple correlation but showed a significant negative correlation by multiple regression controlling the effects of other variables consistently with the hypothesis. On the other hand, Adverse Conditions and Autonomy showed positive and negative correlations in single regression, respectively, but did not show a significant correlation in multiple regression controlling the effects of other variables. Leadership showed no significant correlation in either single regression or multiple regression.

Looking at the results of the three items of Koren & Pető (2020) from the second column of the table, Customer (β=0.595, p<0.01) and Presence (β=0.333, p<0.01) showed a positive correlation, being the same as the single correlation, but Teamwork (β = -0.179, p<0.01) showed a negative correlation. It is theoretically difficult to think that Teamwork has a negative correlation with Physical Proximity, so it can be interpreted as a problem of multicollinearity. This indicates that some caution is required when using the three items of Koren & Pető (2020). By the way, 14 variables that compose the three variables of Koren & Pető (2020) were simply averaged and subtracted from 100 to create one variable Social Distance Index (α=0.727, mean value = 50.753, standard deviation = 9.085) following the way of Crowley & Doran (2020), and examined the relationship with Physical Proximity. Although the problem of multicollinearity was solved, the adjusted R-squared dropped significantly (β=0.437, p<0.01, $R^2$=0.190).

Therefore, in the third column, the results of a regression analysis performed on a variable group in which Customer and Remote Working whose adjusted R-squared for single regression exceed 0.13, the "medium" level (Cohen, 1988), are added to the first column model are shown experimentally. In addition to Customer and Remote Working, four of the five variables shown in the first column, except for the Mechanical Movement, are significant. However, the magnitude of the partial correlation coefficient is different between the first and third columns, and it seems that there is a problem of multicollinearity. This is understandable considering that many of the items that make up Customer and Remote Working were the ones excluded during the exploratory factor analysis due to the load on multiple factors. Despite this problem, the fact that the variables in the first column including Response to Aggression became significant even after controlling the impact of Customer and Remote Working indicates the robustness of the model in the first column. For reference, the adjusted R-squared for all the three models in Table 4, 0.319, 0.293, and 0.491, exceed 0.26, the "large" level (Cohen, 1988). Among them, the value of the mixed model in the third column is the highest. Therefore, although it is worrisome that the distortion of the partial correlation coefficient impairs the accuracy of the influence of individual variables, the mixed model in the third column is considered to be useful for predicting the feasibility of social distances more accurately by taking into account differences in customer contact and outdoor activities.

Table 4 Results of multiple regression analysis with Physical Proximity as the dependent variable

| Variable | β | | | | | |
|---|---|---|---|---|---|---|
| Adverse Conditions | | | - | | - | |
| Leadership | | | - | | - | |
| Information Processing | -0.096 | ** | - | | -0.158 | ** |
| Response to Aggression | 0.460 | ** | - | | 0.155 | ** |
| Mechanical Movement | -0.099 | ** | - | | | |
| Autonomy | | | - | | - | |
| Communication with the Outside | -0.254 | ** | - | | -0.421 | ** |
| Horizontal Teamwork | 0.288 | ** | - | | 0.250 | ** |
| [a]Teamwork | - | | -0.179 | ** | - | |
| [a]Customer | - | | 0.595 | ** | 0.565 | ** |
| [a]Presence | - | | 0.333 | ** | - | |
| [b]Remote Working | | | | | -0.199 | ** |
| $R^2$ | 0.322 | | 0.295 | | 0.494 | |
| Adjusted $R^2$ | 0.319 | | 0.293 | | 0.491 | |
| F | 91.288 | ** | 134.161 | ** | 155.435 | ** |

Note(s): n = 968; *Significance at the 5% level; **Significance at the 1% level; [a] by Koren & Pető (2020); [b] by Dingel & Neiman (2020); ”-“ indicates that it is not used in the regression model. Blank cells indicate not selected in stepwise regression.

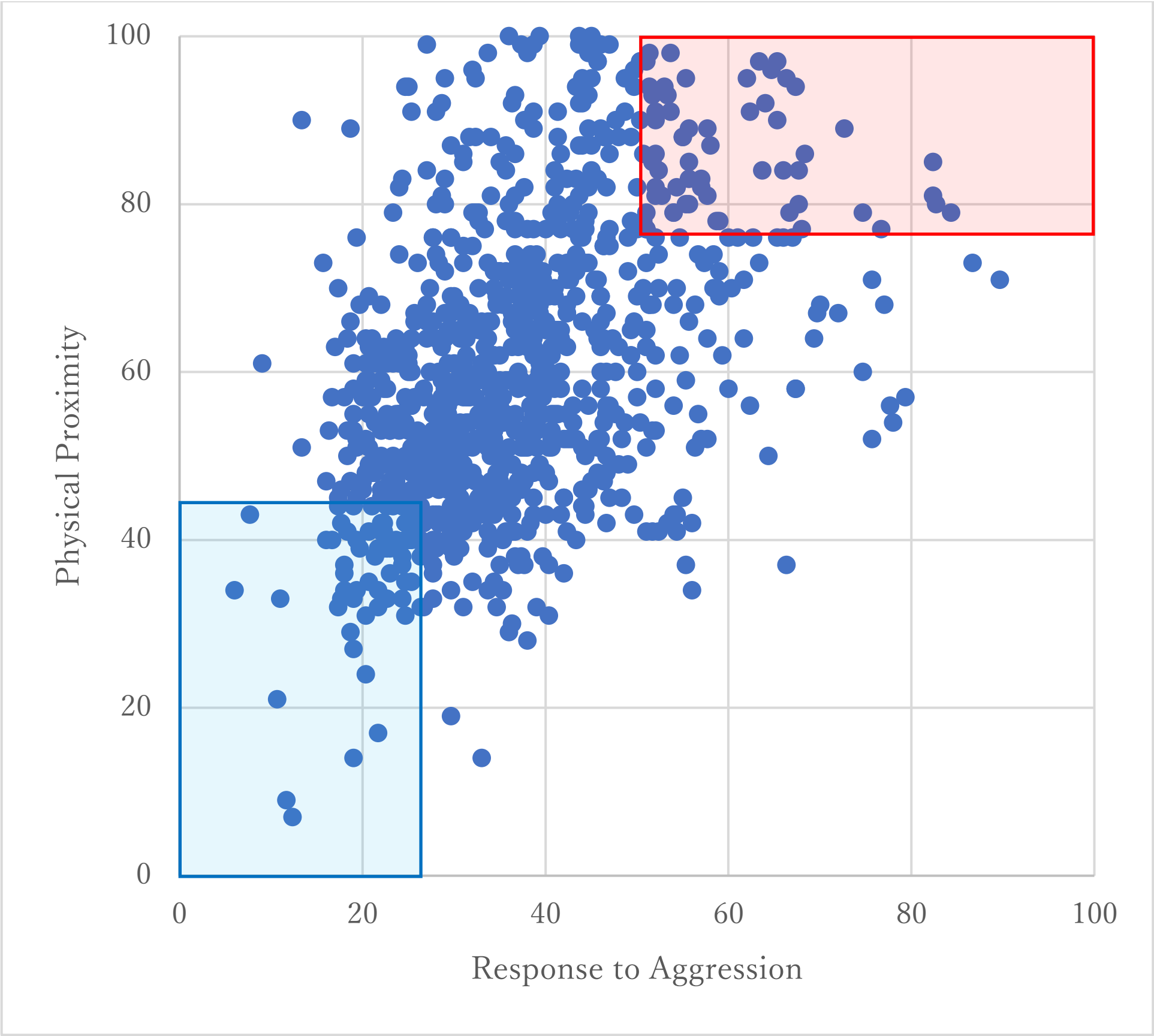

Figure 1 Scatter plot showing the relationship between Response to Aggression and Physical Proximity

Note: The box in the lower left is one standard deviation lower than the average value. The box in the upper right is one standard deviation higher than the average value.

Figure 1 shows the correlation between Response to Aggression and Physical Proximity. The box in the lower left indicates that both Response to Aggression and Physical Proximity values are one standard deviation lower than the average value. The box in the upper right indicates that both Response to Aggression and Physical Proximity values are one standard deviation higher than the average value. Appendix A1 and A2 are extracted from the occupations in the lower left and upper right boxes for each variable. A1 is lined with so-called "experts". On the other hand, in A2, there are teachers of elementary/middle school and special education, therapists, technicians, nurses, employees of restaurants and entertainment facilities, travel/postal service clerks, flight/transportation attendants, etc.

## Discussion

The purpose of this paper is to find variables that correlate with Physical Proximity based on an exploratory method. The scales extracted as a result of the factor analysis were Adverse Conditions, Leadership, Information Processing, Response to Aggression, Mechanical Movement, Autonomy, Communication with the Outside, and Horizontal Teamwork. Of these, Adverse Conditions, Response to Aggression, and Horizontal Teamwork were shown to have a positive correlation with Physical Proximity, and Information Processing, Autonomy, and Communication with the Outside were shown to have a negative correlation with Physical Proximity. Besides, as a result of multiple regression analysis, the model in which five variables of Information Processing, Response to Aggression, Mechanical Movement, Autonomy, Communication with the Outside, and Horizontal Teamwork were simultaneously introduced showed the adjusted R-squared of more than 0.3. Above all, Response to Aggression was shown to be the most important factor affecting Physical Proximity.

In parallel, we verified the relationship between the three variables shown by Koren & Pető (2020) and Physical Proximity and found that the correlation of Customer is significantly larger than that of Teamwork and Presence. It was also shown that the problem of multicollinearity occurred and the Teamwork coefficient was reversed when the three variables are applied to the regression equation at the same time, and it became clear that some caution is required in use. On the other hand, the variables that combine the three variables of Koren & Pető (2020) into one, or the Remote Working Index of Dingel & Neiman (2020) showed a significant correlation with Physical Proximity, avoiding the problem of multicollinearity. However, their adjusted R-squared were below 0.2, which leaves dissatisfaction with the model's explanatory power. In an attempt, we added the Customer of Koren & Pető (2020), which represents the contact with the customer, and the Remote Working of Dingel & Neiman (2020), which represents the small amount of outdoor activity, to our multiple regression model consisting of five variables. As a result, although the partial regression coefficient was distorted due to the problem of multicollinearity, it was significant in four of five variables including Response to Aggression in addition to Customer and Remote Working. This result shows that the model presented in this paper is significant even when the variables in the previous study are controlled. At the same time, it suggests that this mixed model can be used with higher accuracy when predicting the possibility of social distance in the workplace.

In the model shown in this paper, in addition to Response to Aggression, Horizontal Teamwork shows a positive correlation, while Information Processing, Mechanical Movement, and Communication with the Outside show negative correlation, with Physical Proximity. Those who need Response to Aggression or Horizontal Teamwork and have few Information Processing, Mechanical Movement, or Communication with the Outside will have the most difficulty in securing social distance.

## Implication

In the study by Koren & Pető (2020), a model was constructed assuming that one of the three factors that influence social distance is Teamwork. However, the results of this paper show that Response to Aggression has a greater influence on Physical Proximity than on Horizontal Teamwork. This means that while some horizontal teamwork is possible even in the virtual world (Krumm et al., 2016; Painter et al., 2016), it is difficult to respond remotely and on the spot when dealing with people who bring confrontation such as getting angry or offensive. It means that it is necessary to face and deal with them directly. Unfortunately, it seems unlikely that science and technology will replace aggressive human contact. For instance, in the field of robot development, it has been pointed out that it is the most difficult task to operate in an unstructured environment that assumes contact with humans (Mori, 2020).

In this paper, to eliminate arbitrariness as much as possible, we have shown a model for determining physical proximity by executing exploratory factor analysis based on certain rules, extracting eight factors, and performing multiple regression analysis. The types of jobs that are hard to avoid Physical Proximity because of the response to aggression include teachers of elementary/middle school and special education, therapists, technicians, nurses, employees of restaurants and entertainment facilities, travel/postal service clerks, flight/transportation attendants, etc. For such occupations, it may not be easy to secure a social distance by replacing them with remote work, so it is necessary to devise and implement protective measures that can suppress infection even when they are close to people. However, in reality, people whose emotions cannot be suppressed exist in many workplaces not limited to these occupations in the process of contacting customers. For example, the results of an interview survey show that nurses suffer most from aggression from colleagues rather than from patients (Farrell, 1997). Therefore, the results of this paper suggest that the existence of such people may make Physical Proximity, which would otherwise be avoided by remote work, inevitable. Learning how to control own emotions well (Denson et al., 2011), and building a mechanism of reciprocity by increasing the social capital to suppress selfish thinking (Kokubun, 2020; Kokubun et al., 2020), etc. are also considered to be successful in the practice of social distance. At the same time, it should be noted that member aggression often results from mistreatment in the organization. For example, previous studies report that perceived injustice (Beugré, 2005) or unfair treatment of bosses (Neuman and Baron, 1998) increases employee aggression. Therefore, management ingenuity such as devising a method of giving reasonable feedback to the employees should also lead to increasing the possibility of social distance.

## Limitation

This paper exploratively extracted the factors that are the variables used in regression analysis, using the average values by the occupation of the attitude survey data recorded in the US occupation information site, O*NET. Therefore, if the primary data before being aggregated by occupation can be obtained and the results of this paper can be verified, its significance will be great. Besides, Physical Proximity used as the dependent variable of the analysis is a variable based on the questionnaire survey results and may differ from the actual proximity. It is also significant to verify the analytical model in this paper after measuring the actual proximity using GPS location information, etc.

## Conclusion

The spread of new coronavirus (COVID-19) infections continues to increase. The practice of social distance attracts attention as a measure to prevent the spread of infection, but it is difficult for some occupations. Therefore, in previous studies, the scale of factors that determine social distance has been developed. However, it was not clear how to select the items among them, and it seemed to be somewhat arbitrary. In response to this trend, this paper extracted eight scales by performing exploratory factor analysis based on certain rules while eliminating arbitrariness as much as possible. They were Adverse Conditions, Leadership, Information Processing, Response to Aggression, Mechanical Movement, Autonomy, Communication with the Outside, and Horizontal Teamwork. Of these, Adverse Conditions, Response to Aggression, and Horizontal Teamwork had a positive correlation with Physical Proximity, and Information Processing, Mechanical Movement, Autonomy, and Communication with the Outside had a negative correlation with Physical Proximity. Furthermore, as a result of multiple regression analysis, it was shown that Response to Aggression, not the mere teamwork assumed in previous studies, had the greatest influence on Physical Proximity.

## References

Beugré, C. D. (2005). Understanding injustice-related aggression in organizations: A cognitive model. *The International Journal of Human Resource Management, 16*(7), 1120-1136. https://doi.org/10.1080/09585190500143964

Cohen, J. (1988). Statistical power analysis for the behavioral sciences (2nd ed.). Hillsdale, NJ:

Erlbaum.

Cortina, J. M. (1993). What is coefficient alpha? An examination of theory and applications. *Journal of Applied Psychology, 78*(1), 98-104. https://doi.org/10.1037/0021-9010.78.1.98

Crowley, F., & Doran, J. (2020). Covid-19, occupational social distancing and remote working potential in Ireland (No. SRERCWP2020-1). *SRERC Working Paper Series*. Retrieved from https://www.econstor.eu/handle/10419/218897 Accessed 8 Aug 2020.

De Clercq, D., & Belausteguigoitia, I. (2017). Reducing the harmful effect of role ambiguity on turnover intentions. *Personnel Review, 46*(6), 1046-1069. https://doi.org/10.1108/PR-08-2015-0221

den Brock, P., Levy, J., Breckelmans, M. & Wubbels, T. (2005). The Effect of Teacher interpersonal Behavior on Student' Subject specific Motivation. *Journal of Classroom Interaction, 40* (2), 20-33. Retrieved from https://www.jstor.org/stable/23870661 Accessed 8 Aug 2020.

Denson, T. F., Capper, M. M., Oaten, M., Friese, M., & Schofield, T. P. (2011). Self-control training decreases aggression in response to provocation in aggressive individuals. *Journal of Research in Personality, 45*(2), 252-256. https://doi.org/10.1016/j.jrp.2011.02.001

Dingel, J. I., & Neiman, B. (2020). How many jobs can be done at home? Journal of Public Economics, 189. https://doi.org/10.3386/w26948

Farrell, G. A. (1997). Aggression in clinical settings: Nurses' views. *Journal of Advanced Nursing, 25*(3), 501-508. https://doi.org/10.1046/j.1365-2648.1997.1997025501.x

Hatayama, M., Viollaz, M., & Winkler, H. (2020). *Jobs' Amenability to Working from Home: Evidence from Skills Surveys for 53 Countries*. https://doi.org/10.1596/1813-9450-9241

Johns Hopkins University & Medicine (2020). COVID-19 Dashboard. https://coronavirus.jhu.edu/map.html Accessed 8 Aug 2020.

Jong, J. (2016). The role of performance feedback and job autonomy in mitigating the negative effect of role ambiguity on employee satisfaction. *Public Performance & Management Review, 39*(4), 814-834. https://doi.org/10.1080/15309576.2015.1137771 Accessed 8 Aug 2020.

Kokubun, K. (2020). Social capital may mediate the relationship between social distance and COVID-19 prevalence. arXiv preprint arXiv:2007.09939. Retrieved from https://arxiv.org/abs/2007.09939 Accessed 8 Aug 2020.

Kokubun, K., Ino, Y., & Ishimura, K. (2020). Social capital and resilience make an employee cooperate for coronavirus measures and lower his/her turnover intention. arXiv preprint arXiv:2007.07963. Retrieved from https://arxiv.org/abs/2007.07963 Accessed 8 Aug 2020.

Koren, M., & Pető, R. (2020). Business disruptions from social distancing. arXiv preprint arXiv:2003.13983. Retrieved from https://arxiv.org/abs/2003.13983 Accessed 8 Aug 2020.

Krumm, S., Kanthak, J., Hartmann, K., & Hertel, G. (2016). What does it take to be a virtual team player? The knowledge, skills, abilities, and other characteristics required in virtual teams.

*Human Performance, 29*(2), 123-142. https://doi.org/10.1080/08959285.2016.1154061

Mori, N. (2020). Recent status on robot industry in Japan: Focusing on service robots. *Japan Society for Machinery Promotion Economic Research Institute Essay* No.8. Retrieved from http://www.jspmi.or.jp/system/file/6/89/202002essey08_mori.pdf

Neuman, J. H., & Baron, R. A. (1998). Workplace violence and workplace aggression: Evidence concerning specific forms, potential causes, and preferred targets. *Journal of Management, 24*(3), 391-419. https://doi.org/10.1177/014920639802400305

Painter, G., Posey, P., Austrom, D., Tenkasi, R., Barrett, B., & Merck, B. (2016). Sociotechnical systems design: coordination of virtual teamwork in innovation. *Team Performance Management, 22*(7/8), 354-369. https://doi.org/10.1108/TPM-12-2015-0060

Purvanova, R. K., & Bono, J. E. (2009). Transformational leadership in context: Face-to-face and virtual teams. *The Leadership Quarterly, 20*(3), 343-357. https://doi.org/10.1016/j.leaqua.2009.03.004

Stevens, J. P. (1992). *Applied multivariate statistics for the social sciences (2nd edition)*. Hillsdale, NJ: Erlbaum.

Taber, K. S. (2018). The use of Cronbach's alpha when developing and reporting research instruments in science education. *Research in Science Education, 48*(6), 1273-1296. https://doi.org/10.1007/s11165-016-9602-2

van Griethuijsen, R. A., van Eijck, M. W., Haste, H., den Brok, P. J., Skinner, N. C., Mansour, N., Gencer, A. S., & BouJaoude, S. (2015). Global patterns in students' views of science and interest in science. *Research in Science Education, 45*(4), 581-603. https://doi.org/10.1007/s11165-014-9438-6

World Health Organization (2020) Coronavirus disease (COVID-19) advice for the public https://www.who.int/emergencies/diseases/novel-coronavirus-2019/advice-for-public Accessed 8 Aug 2020.

# Appendix

A1 Occupation with low Response to Aggression and low Physical Proximity

| Code | Occupation | Response to Aggression | Physical Proximity |
|---|---|---|---|
| 13-2011.01 | Accountants | 23.0 | 39.0 |
| 15-1111.00 | Computer and Information Research Scientists | 22.7 | 33.0 |
| 15-1131.00 | Computer Programmers | 22.3 | 39.0 |
| 15-1199.03 | Web Administrators | 22.0 | 41.0 |
| 15-1199.08 | Business Intelligence Analysts | 21.7 | 39.0 |
| 15-2011.00 | Actuaries | 20.7 | 35.0 |
| 15-2021.00 | Mathematicians | 16.7 | 40.0 |
| 15-2031.00 | Operations Research Analysts | 18.3 | 41.0 |
| 15-2041.00 | Statisticians | 18.0 | 36.0 |
| 15-2041.01 | Biostatisticians | 19.0 | 33.0 |
| 17-2141.01 | Fuel Cell Engineers | 22.0 | 42.0 |
| 17-2199.10 | Wind Energy Engineers | 21.3 | 38.0 |
| 19-1013.00 | Soil and Plant Scientists | 22.3 | 41.0 |
| 19-1021.00 | Biochemists and Biophysicists | 20.7 | 41.0 |
| 19-1029.01 | Bioinformatics Scientists | 19.3 | 34.0 |
| 19-2011.00 | Astronomers | 20.3 | 31.0 |
| 19-2043.00 | Hydrologists | 19.7 | 39.0 |
| 19-2099.01 | Remote Sensing Scientists and Technologists | 16.0 | 40.0 |
| 19-3011.00 | Economists | 17.7 | 33.0 |
| 19-3011.01 | Environmental Economists | 18.7 | 29.0 |
| 25-1041.00 | Agricultural Sciences Teachers, Postsecondary | 23.7 | 40.0 |
| 25-1125.00 | History Teachers, Postsecondary | 23.0 | 36.0 |
| 27-1012.00 | Craft Artists | 18.0 | 34.0 |
| 27-1013.00 | Fine Artists, Including Painters, Sculptors, and Illustrators | 11.7 | 9.0 |
| 27-1024.00 | Graphic Designers | 21.7 | 34.0 |
| 27-2041.04 | Music Composers and Arrangers | 17.3 | 32.0 |
| 27-3043.05 | Poets, Lyricists and Creative Writers | 19.0 | 14.0 |
| 29-1069.07 | Pathologists | 21.7 | 32.0 |
| 31-9094.00 | Medical Transcriptionists | 17.7 | 42.0 |
| 35-2013.00 | Cooks, Private Household | 10.7 | 21.0 |
| 45-2021.00 | Animal Breeders | 11.0 | 33.0 |
| 45-2091.00 | Agricultural Equipment Operators | 22.7 | 39.0 |
| 45-2092.01 | Nursery Workers | 23.7 | 39.0 |
| 45-2092.02 | Farmworkers and Laborers, Crop | 20.3 | 24.0 |
| 45-3021.00 | Hunters and Trappers | 21.7 | 17.0 |
| 45-4021.00 | Fallers | 12.3 | 7.0 |
| 47-5051.00 | Rock Splitters, Quarry | 7.7 | 43.0 |
| 49-2097.00 | Electronic Home Entertainment Equipment Installers and Repairers | 19.3 | 40.0 |
| 51-4122.00 | Welding, Soldering, and Brazing Machine Setters, Operators, and Tenders | 22.3 | 42.0 |
| 53-7031.00 | Dredge Operators | 19.0 | 27.0 |

## A2 Occupation with many Response to Aggression and high Physical Proximity

| Code | Occupation | Response to Aggression | Physical Proximity |
|---|---|---|---|
| 11-9051.00 | Food Service Managers | 59.0 | 78.0 |
| 21-1093.00 | Social and Human Service Assistants | 66.7 | 79.0 |
| 25-2021.00 | Elementary School Teachers, Except Special Education | 54.0 | 79.0 |
| 25-2023.00 | Career/Technical Education Teachers, Middle School | 58.7 | 78.0 |
| 25-2052.00 | Special Education Teachers, Kindergarten and Elementary School | 57.0 | 83.0 |
| 25-2053.00 | Special Education Teachers, Middle School | 57.7 | 81.0 |
| 25-2059.01 | Adapted Physical Education Specialists | 52.0 | 90.0 |
| 29-1062.00 | Family and General Practitioners | 53.7 | 91.0 |
| 29-1067.00 | Surgeons | 51.0 | 97.0 |
| 29-1069.03 | Hospitalists | 64.0 | 92.0 |
| 29-1071.00 | Physician Assistants | 55.0 | 88.0 |
| 29-1122.00 | Occupational Therapists | 55.3 | 95.0 |
| 29-1124.00 | Radiation Therapists | 51.3 | 98.0 |
| 29-1125.00 | Recreational Therapists | 67.7 | 84.0 |
| 29-1126.00 | Respiratory Therapists | 53.3 | 93.0 |
| 29-1141.00 | Registered Nurses | 53.0 | 94.0 |
| 29-1141.01 | Acute Care Nurses | 66.3 | 95.0 |
| 29-1141.03 | Critical Care Nurses | 67.3 | 94.0 |
| 29-1141.04 | Clinical Nurse Specialists | 55.3 | 80.0 |
| 29-1151.00 | Nurse Anesthetists | 58.0 | 87.0 |
| 29-1161.00 | Nurse Midwives | 50.3 | 97.0 |
| 29-1171.00 | Nurse Practitioners | 55.7 | 85.0 |
| 29-2031.00 | Cardiovascular Technologists and Technicians | 50.3 | 97.0 |
| 29-2033.00 | Nuclear Medicine Technologists | 50.7 | 86.0 |
| 29-2034.00 | Radiologic Technologists | 55.7 | 89.0 |
| 29-2041.00 | Emergency Medical Technicians and Paramedics | 65.3 | 97.0 |
| 29-2052.00 | Pharmacy Technicians | 55.7 | 83.0 |
| 29-2053.00 | Psychiatric Technicians | 82.3 | 85.0 |
| 29-2054.00 | Respiratory Therapy Technicians | 53.7 | 98.0 |
| 29-2061.00 | Licensed Practical and Licensed Vocational Nurses | 65.3 | 90.0 |
| 29-2099.01 | Neurodiagnostic Technologists | 51.3 | 94.0 |
| 29-2099.05 | Ophthalmic Medical Technologists | 49.7 | 94.0 |
| 29-2099.06 | Radiologic Technicians | 49.7 | 96.0 |
| 31-1013.00 | Psychiatric Aides | 72.7 | 89.0 |
| 31-1014.00 | Nursing Assistants | 52.0 | 91.0 |
| 31-1015.00 | Orderlies | 63.3 | 97.0 |
| 31-2012.00 | Occupational Therapy Aides | 62.3 | 91.0 |
| 31-9095.00 | Pharmacy Aides | 50.0 | 82.0 |
| 33-2011.01 | Municipal Firefighters | 57.7 | 89.0 |
| 33-3011.00 | Bailiffs | 82.7 | 80.0 |
| 33-3012.00 | Correctional Officers and Jailers | 82.3 | 81.0 |
| 33-3021.05 | Immigration and Customs Inspectors | 67.7 | 80.0 |
| 33-3051.03 | Sheriffs and Deputy Sheriffs | 84.3 | 79.0 |
| 33-3052.00 | Transit and Railroad Police | 74.7 | 79.0 |
| 33-9031.00 | Gaming Surveillance Officers and Gaming Investigators | 50.7 | 78.0 |
| 33-9093.00 | Transportation Security Screeners | 68.3 | 86.0 |
| 35-1011.00 | Chefs and Head Cooks | 52.7 | 81.0 |
| 35-1012.00 | First-Line Supervisors of Food Preparation and Serving Workers | 62.0 | 95.0 |
| 35-3011.00 | Bartenders | 51.0 | 79.0 |
| 35-9031.00 | Hosts and Hostesses, Restaurant, Lounge, and Coffee Shop | 54.0 | 79.0 |
| 39-1011.00 | Gaming Supervisors | 63.7 | 84.0 |
| 39-1012.00 | Slot Supervisors | 55.7 | 80.0 |
| 39-3011.00 | Gaming Dealers | 55.0 | 88.0 |
| 39-3031.00 | Ushers, Lobby Attendants, and Ticket Takers | 57.0 | 82.0 |
| 39-3091.00 | Amusement and Recreation Attendants | 66.0 | 84.0 |
| 43-4181.00 | Reservation and Transportation Ticket Agents and Travel Clerks | 52.0 | 81.0 |
| 43-5051.00 | Postal Service Clerks | 54.3 | 82.0 |
| 47-2051.00 | Cement Masons and Concrete Finishers | 52.0 | 86.0 |
| 51-3022.00 | Meat, Poultry, and Fish Cutters and Trimmers | 51.7 | 85.0 |
| 53-2021.00 | Air Traffic Controllers | 52.0 | 82.0 |
| 53-2031.00 | Flight Attendants | 64.7 | 96.0 |
| 53-3011.00 | Ambulance Drivers and Attendants, Except Emergency Medical Technicians | 50.3 | 90.0 |
| 53-3022.00 | Bus Drivers, School or Special Client | 52.3 | 84.0 |
| 53-6061.00 | Transportation Attendants, Except Flight Attendants | 51.7 | 93.0 |
| Mean | | 36.7 | 60.3 |
| SD | | 12.9 | 16.9 |
| Mean + SD | | 49.6 | 77.2 |
| Mean - SD | | 23.8 | 43.4 |